# The distinctive symmetry of Bell states.

**Alejandro A. Hnilo**

*Centro de Investigaciones en Laseres y Aplicaciones (CEILAP) (CITEFA-CONICET-UNSAM); Instituto de Investigaciones Científicas y Técnicas para la Defensa (CITEFA), Consejo Nacional de Investigaciones Científicas y Tecnológicas (CONICET), Universidad Nacional de San Martín (UNSAM).*

*Address: CITEFA; J.B. de La Salle 4397, B1603ALO Villa Martelli, Argentina.*
*e-mail: ahnilo@citefa.gov.ar*

*Abstract.*

The Bell's basis is composed of four maximally entangled states of two qubits, named Bell states. They are usual tools in many theoretical studies and experiments. The aim of this paper is to find out the symmetries that determine a Bell state. For this purpose, starting from a general density matrix, physical constraints and symmetry conditions are added until the elements of the Bell's basis are univocally determined. It is found that the usual physical constraints and symmetry conditions do not suffice to determine a Bell state. The additional restriction needed is named here "atomic" symmetry. It is a sort of global symmetry of the system, not derived from the action = reaction law. It is also found that the imperfection in fulfilling the atomic symmetry is linearly proportional to the deviation of the Concurrence from its maximum value. The atomic symmetry allows a different insight on the nature of entanglement, and might be useful as a criterion to define the condition of maximal entanglement for states with more than two qubits.

*August 14th, 2023.*

# 1. Introduction.

Entangled states are an essential resource in Quantum Information. Basic property of entangled states is that correlations measured between their parts are higher than allowed by any classical theory [1,2]. In short, a measurement on one of the parts gives more information about the other part than it is classically allowed. It is sometimes argued that this excess of information comes from the knowledge the observer has of the laws of conservation, or symmetries of the involved Hamiltonian. This is evident in the EPR argument, where the knowledge of the conservation of the total impulse is used to measure momentum and position of two particles (which are emitted in the same momentum-conserving interaction) with arbitrary precision. The Bell's basis naturally describes entangled states of two qubits, which are systems of utmost interest in all Quantum Information and Foundations problems. Therefore, it is pertinent looking for the symmetries, from which that excess of information would come from, for the elements of the Bell's basis.

The answer is not what may be expected at first sight. F.ex., it is customary to say that angular-momentum conservation (which is related with rotational symmetry) determines the fully symmetrical Bell state $|\varphi^+\rangle$. In fact, in the usual notation for particles "a" and "b":

$$|\varphi^+\rangle \equiv (1/\sqrt{2})(|x_a,x_b\rangle+|y_a,y_b\rangle) = (1/\sqrt{2}).(|L_a,R_b\rangle + |R_a,L_b\rangle) \qquad (1)$$

where $|L\rangle$ ($|R\rangle$) is the left (right)-handed circularly polarized state. These states have well defined and opposite angular momentum, so that conservation of angular momentum holds. However, as it will be shown, an additional symmetry is tacitly present in eq.1.

The plan in this paper is as follows: the general form of a two qubits state is written. The well-known physical constraints, and chosen classical symmetries (which are different for each group of Bell states) are applied. At the end of this process there will still be several possible solutions. An additional symmetry is therefore necessary to get a Bell state as the only possible solution. This additional symmetry is found, and commented.

# 2. Derivation of the $|\varphi^+\rangle$ state.

The Bell's basis is $\{|\varphi^\pm\rangle = (1/\sqrt{2})(|x_a,x_b\rangle \pm |y_a,y_b\rangle), |\psi^\pm\rangle = (1/\sqrt{2})(|x_a,y_b\rangle \pm |y_a,x_b\rangle)\}$. The states $|\varphi^+\rangle$ and $|\psi^-\rangle$ are rotationally invariant, and the other two are invariant if the axes are rotated opposite angles in subspaces A and B ("twist" invariance). By the way, an (almost) rotationally invariant basis of the 2-qubits space is $\{|\varphi^+\rangle, |\psi^-\rangle, (1/\sqrt{2})(|\varphi^-\rangle - i |\psi^+\rangle), (1/\sqrt{2})(|\varphi^-\rangle + i |\psi^+\rangle)\}$. The first two elements are properly invariant, having eigenvalues (of the rotation operator) equal to 1. The two last elements have eigenvalues of *modulus* equal to 1.

The matrix of an arbitrary state of polarization of two qubits is, in explicit form:

$$\rho = \begin{pmatrix} Rxxxx & Rxxxy & Rxyxx & Rxyxy \\ Rxxyx & Rxxyy & Rxyyx & Rxyyy \\ Ryxxx & Ryxxy & Ryyxx & Ryyxy \\ Ryxyx & Ryxyy & Ryyyx & Ryyyy \end{pmatrix} \qquad (2)$$

in the basis $\{|x_a,x_b\rangle, |x_a,y_b\rangle, |y_a,x_b\rangle, |y_a,y_b\rangle\}$ from up to down and from left to right. Therefore, the subindexes $(i, j)$ in $Rijkl$ correspond to the subspace A, and the $(k, l)$ ones to the subspace B. Let add now the restrictions:

*1) Features of a physically sounding density matrix.*

   *i)* $\rho$ is self-adjoint

   *ii)* $\text{Tr}(\rho) = 1$

   *iii)* $\rho$ is positive.

The first condition means that $Rikjl=(Rkilj)^*$ and that all the diagonal elements are real numbers. The second one, that the sum of the latter is equal to 1. The third condition is reserved for later use, just to deal with simpler algebra.

*2) Features of the reduced density matrices.*

In a Bell state, single qubits appear unpolarized. This means that each reduced matrix must be equal to one-half the 2x2 identity matrix, or:

$$Rxxxx + Rxxyy = Ryyxx + Ryyyy = \tfrac{1}{2}, \quad Rxxxx + Ryyxx = Rxxyy + Ryyyy = \tfrac{1}{2} \qquad (3a)$$

$$Rxyxx + Rxyyy = Ryxxx + Ryxyy = 0, \quad Rxxxy + Ryyxy = Rxxyx + Ryyyx = 0 \qquad (3b)$$

This reduces the number of independent elements outside the diagonals from four to two.

*3) Naming $a \leftrightarrow b$ and $x \leftrightarrow y$ invariances.*

The naming of the parts in the state is arbitrary, swapping them must leave the state unchanged. It implies that $Rxyyx = Ryxxy = Rxyyx^*$, so that both are real numbers. Also, that $Rxyxx = Rxxxy$, so that all the elements outside the diagonals are determined by a single number. The naming of the polarization axes is also arbitrary. It implies $Rxyxy = Ryxyx = Rxyxy^*$, so that both are real numbers. Also, that $Rxxxy = Ryyyx = -Rxxxy^*$, so that both are pure imaginary numbers.

It is convenient displaying the general form of the matrix at this point:

$$\rho_{\text{aux}} = \begin{pmatrix} d & ig & ig & c \\ -ig & \tfrac{1}{2}-d & f & -ig \\ -ig & f & \tfrac{1}{2}-d & -ig \\ c & ig & ig & d \end{pmatrix} \qquad (4)$$

where: $\{c, d, f, g\}$ are real numbers. Note that states $|\varphi^-\rangle$ and $|\psi^-\rangle$ change sign if restrictions 3 are applied, but that their density matrices remain invariant. Hence, $\rho_{\text{aux}}$ is valid for all Bell states.

*4) Rotational invariance.*

This invariance is distinctive of Bell states $|\varphi^+\rangle$ and $|\psi^-\rangle$. It means that $\mathbf{R}^{-1}\rho_{\text{aux}}\mathbf{R} = \rho_{\text{aux}}$, where $\mathbf{R}$ is the rotation operator (see a helpful expression in the Appendix). It implies that $g = 0 = 4d-2c-2f-1$. Hence, the general form of the rotationally invariant matrix is:

$$\rho_R \equiv \begin{pmatrix} d & 0 & 0 & c \\ 0 & \frac{1}{2}-d & 2d-c-\frac{1}{2} & 0 \\ 0 & 2d-c-\frac{1}{2} & \frac{1}{2}-d & 0 \\ c & 0 & 0 & d \end{pmatrix} \qquad (5)$$

The restriction 1-*iii* (i.e., $\rho$ is positive, which was held in reserve) is easy to handle now. The eigenvalues of $\rho_R$ are: $\{d - c$ (twice), $c - d + \frac{1}{2} \pm |2d - \frac{1}{2}|\}$, and all of them must be positive or zero, hence:

$$\frac{1}{2} \pm |2d - \frac{1}{2}| \geq d - c \geq 0 \qquad (6)$$

All the physical restrictions, plus rotational invariance, have been applied at this point. However, the state is not necessarily entangled. F.ex., in a semi-classical theory of radiation, the source emits pairs of photons with the same well defined polarization, which arbitrarily changes from one pair to the next and is uniformly distributed. The emitted state is described by the density matrix [3]:

$$\rho_{scr} \equiv (1/8) \begin{pmatrix} 3 & 0 & 0 & 1 \\ 0 & 1 & 1 & 0 \\ 0 & 1 & 1 & 0 \\ 1 & 0 & 0 & 3 \end{pmatrix} \qquad (7)$$

which holds to eqs.5 and 6 with $d=3/8$, $c=1/8$. This state does provide some correlation, but not in excess of the classical limits. Here is the point where some extra symmetry must be added to find an entangled state as the only possible solution.

## 3. The "atomic" symmetry.

If "a" and "b" are entangled photons, measuring "a" tells "too much" about "b" because "a" and "b" *are the same thing*. Thinking of a "pair" of photons is essentially erroneous, that's why the term *biphoton* [4] has been coined. The restriction taking into account this feature has not been imposed to the state $\rho_R$ yet. An appropriate mathematical expression for this restriction is found from the following reasoning: as a biphoton is a single entity without internal parts, applying a projector **Q** twice over the subspace A (or B), or instead over subspaces A and B, must lead to the same observable result. This is written:

$$\text{Tr}\,[\rho_R \cdot Q_A(\beta)^t \cdot Q_A(\alpha) \cdot Q_A(\beta)] = \text{Tr}\,[\rho_R \cdot Q_A(\alpha) \cdot Q_B(\beta)] \qquad (8)$$

where $Q_i(j)$ is the operator for a polarizer (the relevant projector here) oriented at angle "j" and placed in the i-subpace (or side) of the Bell's experiment.

In other words: placing a polarizer at an angle $\alpha$ on the A-side and another one at an angle $\beta$

on the B-side (rhs of eq.8) must lead to the same observable result than placing both polarizers on the A-side (lhs). The lhs corresponds to the probability of passage of unpolarized light through two polarizers in the A-side, so that it is given simply the Malus' law. The rhs is calculated using the matrices in the Appendix. Eq.8 then reads:

$$\tfrac{1}{2} \cos^2(\alpha-\beta) = d\,[\cos^2(\alpha-\beta) - \sin^2(\alpha-\beta)] + \tfrac{1}{2}\sin^2(\alpha-\beta) \Rightarrow d = \tfrac{1}{2}. \qquad (9)$$

this result, applied to eq.6, leads to: $0 \geq \tfrac{1}{2} - c \geq 0 \Rightarrow c = \tfrac{1}{2}$. Therefore, the only matrix that holds to all the symmetries is:

$$\tfrac{1}{2}\begin{pmatrix} 1 & 0 & 0 & 1 \\ 0 & 0 & 0 & 0 \\ 0 & 0 & 0 & 0 \\ 1 & 0 & 0 & 1 \end{pmatrix} \qquad (10)$$

which is, as expected, the density matrix of the entangled state $|\varphi^+\rangle$.

I propose for eq.8 the name "atomic" symmetry, because of the original (ancient Greek) meaning of the word "atom" ("not cut" or "without parts"). It is regrettable that we are used to call "atom" a complex ensemble of protons, neutrons and electrons linked by nuclear and electric forces, which is far from being a single object without internal parts. I find curious that Bell states, which actually deserve to be named "atoms", are not necessarily the minuscule objects imagined by Democritus and that we are used to think of, but entities of arbitrary size, that have been observed to be as large as hundreds of kilometers.

Be aware that the atomic symmetry is tacitly present (in addition to the explicit conservation of angular momentum) in the derivation of $|\varphi^+\rangle$ in eq.1, because the chosen combination of $|L\rangle$, $|R\rangle$ is "atomic-symmetric".

### 3. Derivation of the other Bell states.

If the source emits pairs of qubits with crossed polarizations (if "a" is x-polarized then "b" is y-polarized), the polarizer in one station must be rotated an angle $\pi/2$ with respect to the one in the other station to properly take into account the invariance (3) above. Eq.8 must then be written with (f.ex.) $(\alpha+\pi/2)$ instead of $\alpha$ in the rhs. This leads to $d = 0 = c$ and $\rho_R$ becomes the density matrix of the state $|\psi^-\rangle$, which has the form $|xy\rangle$ and is rotationally invariant.

The "twist" symmetry is invariance against a rotation $\theta$ in A-subspace and $(-\theta)$ in B-subspace and is distinctive of Bell states $|\varphi^-\rangle$ and $|\psi^+\rangle$. It replaces rotational invariance (4) above, and means $\mathbf{T^{-1}\rho_{aux}T} = \rho_{aux}$ where $\mathbf{T}$ is the "twist" operator (see a helpful expression in the Appendix). This leads to the general form $\rho_T$ of the twist-invariant matrix:

$$\rho_T = \begin{pmatrix} d & 0 & 0 & c \\ 0 & \tfrac{1}{2}-d & -2d-c+\tfrac{1}{2} & 0 \\ 0 & -2d-c+\tfrac{1}{2} & \tfrac{1}{2}-d & 0 \\ c & 0 & 0 & d \end{pmatrix} \qquad (11)$$

The condition 1-*iii* (positivity) now reads:

$$\tfrac{1}{2} \pm |2d - \tfrac{1}{2}| \geq d + c \geq 0 \qquad (12)$$

But it does not suffice to determine entanglement. Once again, the atomic symmetry must be added. Taking into account the twisted symmetry assumed, eq.8 must be rewritten with ($-\alpha$) instead of $\alpha$ in the rhs (and $\rho_T$ instead of $\rho_R$ of course). This leads to $d = \tfrac{1}{2}$, $c = -\tfrac{1}{2}$ and eq.11 becomes the density matrix of the Bell state $|\varphi^-\rangle$, which has the form $|xx\rangle$ and is twist invariant. If the source emits pairs with crossed polarizations instead, an angle $\pi/2$ must be added as in the case of $|\psi^-\rangle$ to properly take into account the invariance (3), and eq.8 must be written with ($-\alpha+\pi/2$) instead of ($-\alpha$) in the rhs. This leads to $d = 0 = c$ in eq.12, and to the density matrix of the state $|\psi^+\rangle$, which has the form $|xy\rangle$ and is twist invariant.

Therefore, all (fully entangled) Bell states are derived (after the classical symmetries and constraints have been applied) as a consequence of the atomic symmetry. Then: *atomic symmetry* $\Rightarrow$ *full entanglement.* Direct calculation shows that all Bell states hold to the atomic symmetry, then: *full entanglement* $\Rightarrow$ *atomic symmetry.* Finally, all two-qubits states can be written in terms of the Bell's basis. Therefore:

*a two-qubits state is fully entangled $\Leftrightarrow$ it holds to the atomic symmetry*

**4. Imperfect atomic symmetry means imperfect entanglement of the same magnitude.**

To confirm further the equivalence between atomic symmetry and entanglement, let see that the deviations from the perfect case have the same functional dependence on the deviation's parameter. Note that once $\rho_R$ (or $\rho_T$) is obtained, the only effect of the atomic symmetry is to define the value of $d$. Then, the natural way to quantify an imperfection of the atomic symmetry is with a small parameter $\varepsilon$ measuring the deviation from that defined value of $d$. F.ex. for the state $|\varphi^+\rangle$, this means that $d = \tfrac{1}{2} - \varepsilon$. From eq.6:

$$\tfrac{1}{2} - \varepsilon \leq c \leq \tfrac{1}{2} - 3\varepsilon \qquad (13)$$

Using the middle value $c = \tfrac{1}{2} - 2\varepsilon$ and matrix $\rho_R$:

$$\rho_\varepsilon \equiv \begin{pmatrix} \tfrac{1}{2}-\varepsilon & 0 & 0 & \tfrac{1}{2}-2\varepsilon \\ 0 & \varepsilon & \varepsilon & 0 \\ 0 & \varepsilon & \varepsilon & 0 \\ \tfrac{1}{2}-2\varepsilon & 0 & 0 & \tfrac{1}{2}-\varepsilon \end{pmatrix} \qquad (14)$$

Concurrence C(ρ) is a standard measure of entanglement of an arbitrary state ρ. A maximally entangled state has C(ρ) = 1. It is easy to calculate for the case of two qubits [5]:

$$C(\rho_\varepsilon) = 1 - 6\varepsilon \qquad (15)$$

So that the parameter ε, that measures the deviation from perfect atomic symmetry, also measures the deviation from maximum entanglement. A different choosing of the value of *c* in the interval defined by eq.13 leads to a different coefficient of ε in eq.15, being 4 and 8 its limit values. But the precise value of this coefficient is of little interest; what is important here is that both atomic symmetry defect and entanglement defect depend (linearly) of the same parameter.

**5. Summary.**

It is shown that a Bell state matrix is obtained iff the atomic symmetry eq.8 is added to the necessary physical constraints and the usually assumed symmetries. These symmetries are ultimate consequences of the action = reaction law, which applies to an interaction occurring in a single point in space and time (the event of creation of the pair). The atomic symmetry, instead, is a sort of "global" feature, it can involve a non-zero volume of space and time. In my opinion, the nature of the atomic symmetry allows a different view and deeper understanding (than, f.ex., the idea of "nonlocality") of the cause of the excess of correlation displayed by entangled states, and the violation of Bell's inequalities.

It is also shown that an imperfect atomic symmetry means an imperfect entanglement of the same magnitude. It is concluded that "fully entangled state" and "atomic symmetric state" are equivalent expressions, at least for two qubits. Studying if this is also true for the cases of more qubits with different types of entanglement (GHZ, clusters, graphs, etc.) appears interesting and potentially fruitful. For, atomic symmetry may be useful as a criterion to define the state of maximum entanglement in these cases.

**Acknowledgements.**

This work received support from the grants PIP 202200484CO and PUE 229-2018-0100018CO CONICET (Argentina).

**Appendix: Explicit expressions of matrices used in the calculations.**

In the A⊗B space, the rotation matrix an angle θ in A-subspace and an angle φ in B-subspace, for the ordering of the base $\{|x_a,x_b\rangle,|x_a,y_b\rangle,|y_a,x_b\rangle,|y_a,y_b\rangle\}$ as in the main text, is:

$$\begin{pmatrix} \cos\theta.\cos\varphi & \cos\theta.\sin\varphi & \sin\theta.\cos\varphi & \sin\theta.\sin\varphi \\ -\cos\theta.\sin\varphi & \cos\theta.\cos\varphi & -\sin\theta.\sin\varphi & \sin\theta.\cos\varphi \\ -\sin\theta.\cos\varphi & -\sin\theta.\sin\varphi & \cos\theta.\cos\varphi & \cos\theta.\sin\varphi \\ \sin\theta.\sin\varphi & -\sin\theta.\cos\varphi & -\cos\theta.\sin\varphi & \cos\theta.\cos\varphi \end{pmatrix} \quad (A1)$$

To calculate rotational invariance (matrix **R**), make θ = φ. To calculate twist invariance (matrix **T**), make θ = -φ.

In the A⊗B space, a polarizer in the A-subspace oriented at angle α is represented by the matrix $\mathbf{Q_A}(\alpha)$:

$$\begin{pmatrix} \cos^2\alpha & 0 & \cos\alpha.\sin\alpha & 0 \\ 0 & \cos^2\alpha & 0 & \cos\alpha.\sin\alpha \\ \cos\alpha.\sin\alpha & 0 & \sin^2\alpha & 0 \\ 0 & \cos\alpha.\sin\alpha & 0 & \sin^2\alpha \end{pmatrix} \quad (A2)$$

A polarizer in the B-subspace oriented at angle α is represented by the matrix $\mathbf{Q_B}(\alpha)$:

$$\begin{pmatrix} \cos^2\alpha & \cos\alpha.\sin\alpha & 0 & 0 \\ \cos\alpha.\sin\alpha & \sin^2\alpha & 0 & 0 \\ 0 & 0 & \cos^2\alpha & \cos\alpha.\sin\alpha \\ 0 & 0 & \cos\alpha.\sin\alpha & \sin^2\alpha \end{pmatrix} \quad (A3)$$